\documentclass[11pt]{article}
\begin{document}


\begin{flushright}
\hfill{UPR-918-T}\\
\hfill{HU-EP-00/63}\\
\hfill{IHP-2000/13}\\
\hfill{hep-th/0101007}\\   
\hfill{December 2000}\\
\end{flushright}

\vspace{15pt}

\begin{center}{ \Large{\bf Gauging of N=2 Supergravity
Hypermultiplet \\ and 
Novel Renormalization Group Flows}}

\vspace{15pt}

{Klaus Behrndt$^a$\footnote{E-mail:behrndt@physik.hu-berlin.de} 
 and Mirjam Cveti\v c$^b$ \footnote{E-mail:cvetic@cvetic.hep.upenn.edu}
 }
\vspace{15pt}

{\it $^a$ Humboldt Universit\"at zu Berlin,
Institut f\"ur Physik,\\ 
Invalidenstrasse 110, 10115 Berlin,
 Germany\\
$^b$ Department of Physics and Astronomy, \\
University of Pennsylvania, Philadelphia PA 19104-6396, USA\\ 
and\\
Institut Henri Poincar\'e, \\
11 rue Pierre et Marie Curie, F75231 Paris, Cedex 05, France\\
}
\vspace{40pt}   
  
\underline{ABSTRACT}
\end{center} 

We provide the explicit gauging of all the $SU(2,1)$ isometries of one
N=2 supergravity hypermultiplet, which spans $SU(2,1)/U(2)$ coset
space parameterized in terms of two complex projective coordinate
fields $z_1$ and $z_2$.  We derive the full, explicit Killing
prepotential that specifies the most general superpotential.  As an
application we consider the supersymmetric flow (renormalization
group) equations for:
(i) the flow from a null singularity to the flat, supersymmetric
space-time and (ii) the flow that violates c-theorem with the
superpotential crossing zero.
   
\newpage
\newcommand{\be}[3]{\begin{equation}  \label{#1#2#3}}
\newcommand{\bib}[3]{\bibitem{#1#2#3}}
\newcommand{\ee}{\end{equation}}
\newcommand{\ba}{\begin{array}}
\newcommand{\ea}{\end{array}}
\newcommand{\p}{\partial}
\renewcommand{\arraystretch}{1.5}

\let\huge=\Large
\let\LARGE=\Large
\let\Large=\large



\section{Introduction}

BPS domain wall configurations in five-dimensional $N=2$ gauged
supergravity provide a fertile ground to address (see,
\cite{bc1,kls,kl,bc2} and references therein) in fundamental theory
the candidate solutions for trapping of gravity \cite{rs2} and within
AdS/CFT correspondence \cite{m} for the study of viable, non-singular,  
gravity duals of four-dimensional strongly coupled $N=1$ supersymmetric
field theories (see, \cite{fgpw, ks,mn} and references therein).
   
On the other hand, the gauging procedure of the five-dimensional
supergravity has been rather poorly understood until recently.  While
the Abelian $U(1)_R$-gauging with the vector supermultiplets, only,
was known for a while \cite{gst}, the progress on non-Abelian gauging
of vector and tensor multiplets was made only recently \cite{gz}.  In
addition, the most recent efforts are focused on the gauging of
hypermultiplets \cite{an/be,lu/ov,cd,bhl}.  Nevertheless the full
fledged gauging with even single hypermultiplet has not been done,
yet.

The purpose of this paper is twofold.  First, we provide an explicit
gauging of one-hypermultiplet superfield of N=2 supergravity which
spans the $SU(2,1)/U(2)$ coset space; we choose to parameterize in
terms of projective coordinate fields $z_1$ and $z_2$.  We find the
explicit prepotential that specifies the most general
superpotential. The details are given in Section 2, and the explicit
form of the eight prepotentials and the resulting superpotential is
displayed in the Appendix.

Second, we employ this newly obtained theory to study novel domain
wall solutions (which specify novel renormalization group (RG) flows
of dual field theories).  In particular, we focus on the potential,
which involves the scalar fields of the hypermultiplets, only.  In
Section 3 we study examples of supersymmetric flows in the case of the
Abelian gauging associated with the Killing directions in the subset
of compact directions.  In particular, we quantitatively analyze
the flow from a null singularity to the flat space-time, reminiscent
of the solution in \cite{mn}, and another flow that violates the
c-theorem and interpolates between ``infra-red'' anti-deSitter space
time and null-singularity which is reminiscent of the solution in
\cite{clp2}. This latter solution was found as a geodesic extension
the D3-brane configuration (behind the horizon) that is further
compactified on a five-sphere; it was proposed as a possible candidate
for a gravity trapping domain wall.  Possible further applications and
concluding remarks are given in Section 4.

While this work was in progress we were informed of a related work in
progress \cite{cdvk}, where hypermultiplet gauging has been pursued in
a basis where the axionic $U(1)$ symmetry is manifest.


\section{Gauging  the Isometries of a Hypermultiplet}


The focus of this paper will be on the scalar fields of the
hyper-supermultiplet, only. Thus, as a concrete application we shall
concentrate on the Abelian gauging of the most general hypermultiplet
isometries. Of course, since we provide the explicit prepotential for
the complete hypermultiplet isometry, generalizations to the 
non-Abelian gauging, that also involves vector supermultiplets is
straightforward,   and will
be studied, along with applications to  the RG flows, elsewhere
\cite{bcp}.

 We parameterize\footnote{Another 
possible parameterization involves the complex
fields $S$ and $C$, where, e.g., the action of the axionic $U(1)$
symmetry is manifest, see \cite{fesa}.  
There the K\"ahler potential is of the form $K
= -\log( S + \bar S - 2 C\bar C)$ and is related to that in
Eq. (\ref{160}) by the K\"ahler transformation combined with the
reparameterization: $ z_1= (1-S)/(1+S)$ and $z_2=2C/(1+S)$. The
gauging procedure in either parameterization is expected to give
equivalent results.}  
the coset space $SU(2,1)/U(2)$ of the universal
hypermultiplet with two complex scalars $z_1$ and $z_2$ with the
K\"ahler potential:
\be160
K = - \log(1- |z_1|^2 - |z_2|^2 ), 
\ee
with $|z_1|^2 +|z_2|^2 < 1$. The K\"ahler metric and the K\"ahler 
two-form take the form:
\be170
\ba{rcl}
\partial_A \partial_{\bar B} K\, dz^A d\bar z^B 
&=& e^{K} \delta_{AB} \, dz^A d\bar z^B + e^{2K}
(\bar z_A dz^A)(z_B d\bar z^B)\, ,  \\
\partial_{A}\partial_{\bar B} K\,  dz^A \wedge d\bar z^B
&=& e^{K} \delta_{AB} \, dz^A \wedge d\bar z^B+ e^{2K}
 (\bar z_A \, dz^A) \wedge (z_B d\bar z^B) \ .
\ea
\ee
In the following Subsection we shall first focus on the quaternionic
structure and the isometries. In the subsequent Subsections
we provide the gauging of the isometries, determine the prepotentials
associated with all the isometries (which are summarized in the Appendix)
and provide the explicit form of the superpotential for the specific
examples of Abelian gauging.  These latter examples are then employed
in Section 3 to illustrate novel RG flows.


\subsection{The Quaternionic Structure and the Isometries}


Let us start with a discussion of the quaternionic structure of this
space.  Following essentially the parameterization employed in
\cite{bsv}, it turns out to be more convenient to introduce polar
coordinates in the following way:
\be190
z_1 = r \, (\cos\theta/2) \, e^{i (\psi + \varphi)/2} \qquad , \qquad
z_2 = r \, (\sin\theta/2) \, e^{i (\psi - \varphi)/2}\, .
\ee
with $r \in [0,1),\ \theta \in [0, \pi),\ \varphi \in [0, 2 \pi)$
and $\psi \in [0, 4\pi)$. The K\"ahler metric then becomes:
\be200
\partial_A \partial_{\bar B} K\, dz^A d\bar z^B 
= {dr^2 \over (1- r^2 )^2} + {r^2 \over 4 (1- r^2 )}
(\sigma_1^2 + \sigma_2^2 ) +{r^2 \over 4 (1- r^2 )^2} \sigma_3^2 \ ,
\ee
where the $SU(2)$ one-forms ($d\sigma_i +{1\over 2}
\epsilon_{ijk} \sigma_j \wedge \sigma_k =0$) are given by:
\be210
\ba{l}
\sigma_1 = \cos\psi \, d\theta + \sin\psi \, \sin\theta\, d\varphi \ , \\
\sigma_2 = - \sin\psi \, d\theta + \cos\psi \, \sin\theta \, d\varphi \ , \\
\sigma_3 = d\psi + \cos\theta \, d\varphi  \ .
\ea
\ee
In terms of these one-forms we find  the following expressions
for the Vielbeine:
\be220
e^r = { dr \over 1 -r^2} \quad , \quad 
e^3 = {r \over 2(1-r^2)} \sigma_3 \quad , \quad
e^{1/2} = {r \over 2 \sqrt{1 - r^2}} \sigma_{1/2} \,,
\ee
which in the  complex notation take the form: 
\be224
v = {1 \over 1 -r^2}(dr + i \, {r \over 2} \, \sigma_3)
\qquad , \qquad u = -{r \over 2\sqrt{1-r^2}} (\sigma_2 + i \, \sigma_1)\ .
\ee
The metric is then of the following form:
\be226
ds^2 = e^r e^r + e^1 e^1 + e^2 e^2 + e^3 e^3
= u \bar u + v \bar v\, .
\ee
Since this space is quaternionic, it allows for a triplet of the complex
structure $J^i_{mn}$, giving rise to a triplet of K\"ahler two-forms:
$\Omega^i = e^m J^i_{mn} \wedge e^n$, which can be written as:
\be230
\ba{l}
\Omega^1 = {r \over (1-r^2)^{3/2}} \Big[ dr \wedge \sigma_1
+ {r \over 2} \sigma_2 \wedge \sigma_3 \Big]\, ,  \\
\Omega^2 = {r \over (1-r^2)^{3/2}} \Big[ - dr \wedge \sigma_2
+ {r \over 2} \sigma_1 \wedge \sigma_3 \Big] \, , \\
\Omega^3 = {r \over (1-r^2)^{2}} dr \wedge \sigma_3
+ {r^2 \over 2(1-r^2)} \sigma_1 \wedge \sigma_2   \ .
\ea
\ee
The holonomy group of a quaternionic space is contained in $SU(2) \times
SP(2m)$ and the K\" ahler two-forms have to be covariantly constant with
respect to the $SU(2)$ connection $p^i$: $\nabla \Omega^i = d \Omega^i
+ \epsilon^{ijk} p^j \wedge \Omega^k = 0$, i.e. $\Omega^i$ preserves
the quaternionic algebra. For our specific case the $SU(2)$
connections are:
\be250
p^1 = - {\sigma_1 \over \sqrt{1-r^2}} \quad , \quad 
p^2 =  {\sigma_2 \over \sqrt{1-r^2}} \quad , \quad 
p^3 = -{1 \over 2}(1 + {1 \over 1-r^2}) \, \sigma_3\,,
\ee
and fulfill the following relationship:
\be240
d p^i + {1 \over 2} \, \epsilon^{ijk} p^j \wedge p^k = - \Omega^i \ .
\ee

The isometry group of this space is $SU(2,1)$ whose the eight generators
are specified by the following eight Killing vectors, see also 
\cite{bsv}:
\be182
\ba{ll}
k_1 = {1 \over 2i} \Big[ z_2 \partial_{z_1} + z_1 \partial_{z_2} - c.c. \, 
	\Big] \ , &
k_2 = {1 \over 2} \Big[ -z_2 \partial_{z_1} + z_1 \partial_{z_2} + c.c. 
	\, \Big] \ , \\
k_3 = {1 \over 2i} \Big[ -z_1 \partial_{z_1} + z_2 \partial_{z_2} - c.c. \,
	\Big] \ , &
k_4 = {1 \over 2i} \Big[ z_1 \partial_{z_1} + z_2 \partial_{z_2} - c.c. \,
	\Big] \ , \\
k_5 = {1 \over 2} \Big[(-1 +z_1^2) \partial_{z_1} + z_1 z_2 \partial_{z_2}
	+c.c. \Big] \ , &
k_6 = {i \over 2} \Big[(1 +z_1^2) \partial_{z_1} + z_1 z_2 \partial_{z_2}
	-c.c. \Big] \ , \\
k_7 = {1 \over 2} \Big[-z_1z_2 \partial_{z_1} + (1-z_2^2) \partial_{z_2}
	+c.c. \Big] \ , &
k_8 = {i \over 2} \Big[z_1 z_2 \partial_{z_1} + (1+z_2^2) \partial_{z_2}
	-c.c. \Big] \ .
\ea
\ee
The compact subgroup $SU(2)\times U(1)$ is associated with the Killing
vectors $(k_1,\cdots,k_4)$ and the non-compact isometries are
parameterized by $(k_5,\cdots,k_8)$.  The two Abelian isometries are
the phase transformations of $z_1$ and $z_2$ and correspond to the
Killing vectors $k_3$ and $k_4$, respectively.  In addition, the
action of the $SU(2)$ subgroup corresponds to the ``rotations'' of the
two complex coordinates $z_{1,2}$ and the three generators, which are
represented by $(k_1, k_2, k_3)$, fulfill the $SU(2)$ algebra $[k_m
,k_n] = i \epsilon_{mnp} k_p$.


\subsection{Gauging of Abelian Isometries}


We gauge only a single combination of the isometries $k =
a^n k_n$ associated with the graviphoton $A$, i.e.\ the covariant
derivative of the hyper scalar $q^u$ becomes: $dq^u \rightarrow dq^u +
k^u A$, where $u$ is an index of the quaternionic manifold.
Supersymmetry with eight unbroken supercharges requires that the
Killing vector has to be tri-holomorphic. (For details see
\cite{an/be,cd} and references therein.)  This property is ensured if
the Killing vector $k_n$ can be expressed in term of a Killing
prepotential $P_n^i$:
\be280
(\Omega^i \cdot k_n) = -d P^i_n - \epsilon^{ijk} p^j P^k_n \, ,
\ee
where ``$i$'' is the $SU(2)$ index and the K\"ahler forms are defined
in (\ref{230}). We indeed derive the Killing prepotentials associated
with all eight Killing vectors in (\ref{182}).  Their explicit form is
given in the Appendix. 

In addition, there is a further constraint coming from the fact, that
the fermionic projector has to commute with the covariant
derivative. This constraint, which was discussed as geodesic
constraint in N=1,D=4 supergravity \cite{cgr} and which we will
discuss in more detail in \cite{bcp}, becomes for the case at hand:
$dq^u\, [P_n , \nabla_u P_m] \, a^n a^m =0$ and reads in components
\be193
\epsilon^{ijl} \, P^j_n \, \Omega_{uv}^l \, dq^u k^v a^n = 0 \ .
\ee
This constraint (\ref{193}) puts severe restrictions on consistent
superpotentials and especially seems to exclude a regular
flow\footnote{In the previous version of this paper we derived a
regular flow by considering Abelian gauging of compact
isometries. However we did not take into account this constraint, 
which is not satisfied for the considered regular flow.}.

As a concrete example let us first start with a gauging of a linear
combination of the two Abelian Killing vectors, only:
\be271
k = a^3 \, k_3 + a^4\, k_4 \ .
\ee
where the special case 
$a^3+a^4=0$ was already discussed in \cite{bhl}.  
The (real-valued) superpotential $W$, given by the
determinant of the $SU(2)$-valued Killing prepotential \cite{bhl},
takes for the above Killing vector (\ref{271}) the form:
\be310
\ba{l}
W^2 = \det( - {\cal P} ) = 
-(P^1)^2 - (P^2)^2 - (P^3)^2 = \\
= { 1 \over 4 (1-r^2)^2}
\Big[ 4 (a^3)^2 (1 -r^2) \sin\theta^2  + \Big( a^3 (2-r^2) \cos\theta +a^4 
r^2
\Big)^2 \Big]\, . 
\ea
\ee
Here ${\cal P}$ is the matrix-valued Killing prepotential, i.e.
${\cal P}\equiv P^i \tau_i$, where $\tau_i$ are Pauli matrices
$(i=1,2,3)$. This potential is consistent with (\ref{193}) at the
critical values: $\theta =0, \pi$ and has one non-trivial fixed point
at $r= 0$ (and for any value of $\theta$).  At this fixed point only
the $r$-direction is non-flat and we have
\be311
(\partial^2_r \log |W|)_0 = (1 + {a^4 \over a^3} \cos\theta) \ .
\ee
If this expression is positive it is an UV attractive fixed point 
and an IR fixed point if it is negative. In both cases the flow goes
towards a singularity, but in the latter case $W$ passes a zero at
$\{\theta = 0 , r^2 = {2 \over {1 - a^4/a^3}}\}$.  One can restrict
oneself to the critical orbit with $\theta =0$, where the
superpotential becomes:
\be342
W = { {a^3(2-r^2) + a^4r^2} \over {2(1-r^2)}} = a^3 + {(a^3 + a^4) \, r^2 
\over  2\, (1-r^2)} \, .
\ee

Next, we are going to discuss  modifications if we gauge also the
other compact Killing isometries, i.e.\ we consider the Killing vector:
\be624
k = a^1 k_1  + a^2 k_2 + a^3 k_3 + a^4 k_4 \ .
\ee
The superpotential $W$ takes the form:
\be340
W^2 = \Big[ (P^1_n a^n)^2 + (P^2_n a^n)^2 + (P^3_n a^n)^2 
	\Big]\ ,
\ee
where the explicit form of the prepotentials $P^i_n$ is given in the
Appendix. In this case the constraint (\ref{193}) implies non-trivial
restrictions on the coefficients and consistent cases are: (i) either
$a^1=a^2=0$ yielding the case discussed before or (ii) $a^3=a^4=0$
combined with $\theta = d \varphi = d\psi =0$,
which yields the superpotential:
\be274
W^2 = {(a^1)^2 + (a^2)^2 \over 
	1-r^2} \ .
\ee


\section{Application: Novel Renormalization Group Flows}


The results of the previous Section and Appendix  in principle allow for
the full analysis of the supersymmetric extrema of the general potential
as well as the study supersymmetric  flows between such isolated
extrema. However, due to the complexity of the potential we confine
ourselves to special cases. 
 
In a general case the linear combination of these isometries is
weighed with the constant coefficients:
\be410
{\bf a}= (a^1,a^2,a^3,a^4,a^5,a^6,a^7,a^8)\, .
\ee

We analyze special cases that involve only isometries in the Cartan
subalgebra, i.e. $a^3,a^4 \neq 0$.  The corresponding superpotential
was given in Section 2.2 (\ref{342}). A novel feature is that now the
superpotential can pass a zero and in a special case this point can
even be extremal, i.e.\ $W =dW =0$. Namely, the zeros of the
superpotential (\ref{342}) take place at:
\be520
r\equiv r_0=\sqrt{ 2 \over(1-\delta)}\ , 
\ee
where $\delta = {a^4 \over a^3}$ and thus the only value of $\delta$, for
which $r_0\le 1$, is $\delta \le -1$.  It is an extremum if 
$\delta \rightarrow - \infty$, i.e.\ $a^3=0$, $a^4 \neq 0$
(which is obvious from the superpotential (\ref{342})). As pointed out
at the end of Subsection 2.2, in this case we have only one extremum,
which can be in the UV or IR regime (see eq.\ (\ref{311})).  In the IR
case, the superpotential necessarily passes zero along the flow.

In order to solve the flow equation, we introduce a coordinate
system, where
\be834
ds^2 = e^{2A} \Big(-dt^2 + d\vec x^2 \Big) + dy^2\, ,
\ee
and the flow equations become \cite{bc1}:
\be635
\partial_y A = - W \qquad , \qquad \partial_y \, r = 3 g^{rr} \partial_r W\,
,
\ee
and the solution for the superpotential (\ref{342}) is
\be736
r = e^{3 (a^3 + a^4) (y-y_0)} \qquad , \qquad e^{2A} = e^{-2 a^3 (y-y_0)}
	\sqrt{1 - e^{6 (a^3 + a^4) (y-y_0)}}\, .
\ee
Thus $W=dW=0$ point is reached for the special case: $a^3=0$ and $a^4
y \rightarrow -\infty$ ($\delta=-\infty$).  This is a
special example where the UV point is singular while the IR regime
corresponds to the flat (supersymmetric) space-time. The solution
resembles that of \cite{mn}, where the UV region is formally singular
(since it corresponds to a decompactification to a ``dilatonic'' D=7
space-time), while in the IR it becomes a flat D=5 space-time.  That
type of solutions may provide useful supergravity duals for testing
the IR behavior of N=1 supersymmetric field theories.

On the other hand, $a^3 + a^4<0$ and $\delta<-1$
corresponds to the flow that in the IR regime passes $W=0$ and runs
into the null-singularity with $W \rightarrow - \infty$.  This set of
solutions is intriguing since it violates the c-theorem. It bears
similarities with the solution in \cite{clp2} describing the inside
horizon region of D3-brane, that is subsequently compactified on a
five-sphere.  In the latter case the singularity is, however, naked;
if one were able to to identify a (stringy) mechanism to regulate this
singularity such a domain wall solution could trap gravity.

Finally, let us mention that it is not enough to focus only on the
superpotential. For example, consider the Killing vector
\be284
k = a^3 k_3 + a^4 k_4+ a^5 k_5 
\ee
where $k_5$ is a non-compact Killing vector.  One finds that
$\partial_{\theta}W =0$ and the constraint (\ref{193}) is satisfied
if $\theta = 0$ yielding the superpotential:
\be302
W = {a^3 \, (r^2 -2)   + a^4\, r^2 + 
2 \, a^5 \, r \, \sin\alpha \over 2 \, (1-r^2)} \ ,
\ee
with $\alpha = {1 \over 2} (\varphi + \psi)$. 
This superpotential has two extrema:
\be462
r_+=0 \ , \ \alpha=0 \ , \qquad {\rm and} \qquad r_-= 
{\gamma \over 2} - \sqrt{ {\gamma^2 \over 4}  - 1}  \ , \ 
\alpha ={\pi \over 2} \ ,
\ee
with $\gamma = {a^3 -a^4 \over a^5}>2$.  However, the first extremum is
{\em not} a fixed point of the $\alpha$-flow, because the metric
component $g^{\alpha \alpha}$ has a pole at this point and one does not
obtain an AdS vacuum. How about the cases discussed before, is the
extremum at $r=0$, $\theta=0$ perhaps also an artifact of the coordinate
system? The explicit solution (\ref{736}) already shows that there is
good AdS vacuum, but one may also go back to the $z_{1/2}$ coordinates
with the metric given in (\ref{170}). Using the relations $r^2 = z_1 \bar
z_1 +z_2 \bar z_2$ and $r^2 \cos\theta = z_1 \bar z_1 -z_2 \bar z_2$
it is straightforward to transform the superpotential (\ref{310}) in the
$z_{1/2}$ coordinates and the extremum at $r=\theta=0$ translates into
$z_1 = z_2 = 0$, which is a regular point for the corresponding metric
(\ref{170}).


\section{Concluding Remarks and Open Avenues}


In this paper we have provided the gauging of the full $SU(2,1)$
isometry group of the universal hypermultiplet spanning the
$SU(2,1)/U(2)$ coset space. We have chosen the parameterization in
terms of the complex projective space fields $z_1$ and $z_2$ and
determined the full Killing prepotential.

We analyzed two sets of flows associated with the gauging of the
Cartan subalgebra.  It corresponds to the gauging of the single $U(1)$
and involves the flow from the UV singular point to the IR
supersymmetric flat spacetime ($dW=W=0$). Another flow, which
corresponds to the gauging of both isometries in the Cartan
subalgebra, provides a flow that violated the c-theorem.

Since we have derived the general prepotentials associated with the gauging
of the full isometry group, one can now proceed with the general
non-Abelian gauging  which necessarily involves vector-multiplets as well. 
In this case the structure of the scalar potential is significantly more
complicated and its analysis is deferred for further study \cite{bcp}. 

The results here provide a stepping stone toward a general procedure
to gauge an arbitrary number of hypermultiplets and the subsequent
analysis of the vacuum structure for such general N=2 gauged supergravity
theories that awaits further study.


\vskip 0.5cm
\noindent{ {\bf Acknowledgments} We would like to thank Gianguido
Dall'Agata for many enlightening discussions and suggestions and
Anna Ceresole for pointing out typographical errors in the appendix
of the original version. We would
also like to thank him for informing us about the related work in
progress \cite{cdvk}. 
M.C. would like to thank the Theoretical
Particle Physics Group of the Humboldt University  and the Center for 
Applied Mathematics and Theoretical Physics, Maribor,
Slovenia for hospitality during the final stages of the project.  The work
is supported in part  by
a DFG Heisenberg Fellowship (K.B.), the U.S. Department of Energy Grant
No.  DOE-EY-76-02-3071 (M.C.), the NATO Linkage grant No. 97061
(M.C.) and  the programme {\it Supergravity, Superstrings and
M-theory} of the Centre \'Emile Borel of the Institut Henri Poincar\'e
No. UMS-839-CNRS/UPMC (M.C.).}

\newpage

\appendix{\bf \Large Appendix: Killing Prepotentials}

\bigskip


The Killing prepotential associated with gauging of the isometries  in the
compact subgroup $SU(2)\times U(1)$, specified by the Killing vectors
$(k_1,\cdots,k_4)$ defined in (\ref{182}):

\renewcommand{\arraystretch}{1.2}
\be600
\ba{l}
P_1 = {1 \over \sqrt{1- r^2}} \left(\ba{c} 
\cos\psi \sin\varphi +\cos\theta \sin\psi \cos\varphi\\
\sin\psi \sin\varphi -\cos\theta \cos\psi \cos\varphi\\
- {2 - r^2\over 2\sqrt{1 - r^2} }  \sin\theta \cos\varphi 
\ea \right)\, ,  \\[15mm]
P_2 = {1 \over \sqrt{1- r^2}} \left(\ba{c} 
\cos\psi \cos\varphi -\cos\theta \sin\psi \sin\varphi\\
\sin\psi \cos\varphi +\cos\theta \cos\psi \sin\varphi\\
{2 - r^2\over 2\sqrt{1 - r^2} } \sin\theta \sin\varphi 
\ea \right)\, ,  \\[15mm]
P_3 = {1 \over \sqrt{1- r^2}} \left(\ba{c} 
\sin\psi \sin\theta \\
-\cos\psi \sin\theta\\
{2 - r^2\over 2\sqrt{1 - r^2} } \cos\theta 
\ea \right) 
\quad , \quad
P_4 = - {r^2 \over 2(1- r^2)} \left(\ba{c} 0 \\
0  \\
1
\ea \right)\, . 
\ea
\ee

The prepotentials associated with the non-compact isometries, specified by
the Killing vectors $(k_5,\cdots, k_8)$ defined in (\ref{182}), are of the
form:

\renewcommand{\arraystretch}{1.2}
\be610
\ba{l}
P_5 = -{r \over 1- r^2} \left(\ba{c} 
\sqrt{1-r^2} \sin{\theta \over 2} \cos{\varphi - \psi \over 2} \\
-\sqrt{1-r^2} \sin{\theta \over 2} \sin{\varphi - \psi \over 2}  \\
\cos{\theta \over 2} \sin{\varphi + \psi \over 2} 
\ea \right)
\quad , \quad
P_6 = {r \over 1- r^2} \left(\ba{c} 
\sqrt{1-r^2} \sin{\theta \over 2} \sin{\varphi - \psi \over 2} \\
\sqrt{1-r^2} \sin{\theta \over 2} \cos{\varphi - \psi \over 2}  \\
-  \cos{\theta \over 2} \cos{\varphi + \psi \over 2} 
\ea \right)\, ,  \\[15mm]
P_7 = -{r \over 1- r^2} \left(\ba{c} 
\sqrt{1-r^2} \cos{\theta \over 2} \cos{\varphi + \psi \over 2} \\
\sqrt{1-r^2} \cos{\theta \over 2} \sin{\varphi + \psi \over 2}  \\
 \sin{\theta \over 2} \sin{\varphi - \psi \over 2} 
\ea \right)
\quad ,\quad
P_8 = {r \over 1- r^2} \left(\ba{c} 
\sqrt{1-r^2} \cos{\theta \over 2} \sin{\varphi + \psi \over 2} \\
-\sqrt{1-r^2} \cos{\theta \over 2} \cos{\varphi + \psi \over 2}  \\
- \sin{\theta \over 2} \cos{\varphi - \psi \over 2} 
\ea \right)\, . \\
\ea
\ee


\end{document}